\documentclass[12pt]{article}
\usepackage{graphicx,fullpage}
\begin{document}
\title{Localization of light in a lamellar
  structure with left-handed medium : the light wheel}
\author{P. H. Tichit, A. Moreau and G. Granet\\
LASMEA, UMR CNRS 6602, Universit\'e Blaise Pascal,\\
24 avenue des Landais, 63177 Aubi\`ere, France.}
\date{}
\maketitle

\begin{abstract}
The contra-directional coupling between a left-handed monomode
waveguide and a right-handed monomode waveguide is rigorously
studied using a complex plane analysis. Light is
shown to rotate in this lamellar structure 
forming a very exotic mode which we have called a light
wheel. The light wheel can be excited using evanescent coupling
or by placing sources in one of the waveguides.
This structure can thus be seen as a new type of
cavity. It is a way to suppress the guided mode
of a dielectric slab.
\end{abstract}

\begin{twocolumn}
Left-Handed Materials (LHM) present simultaneously negative
permittivity and permeability. Such materials
were a pure theoretical oddity\cite{veselago67}
until the recent experimental demonstration of the negative
refraction\cite{shelby01}. Left-handed materials can be made using
metamaterials, {\em i.e.} periodical structures 
with a period smaller than the wavelength, which behave as
a homogeneous medium. Consequently to these works the LHM have aroused
huge interest. It clearly appeared that LHM had
hardly been studied although they could lead to structures 
presenting highly nonconventional behaviours : they seemingly allow to
overcome the Rayleigh limit\cite{pendry00} and could even lead to
invisibility\cite{pendry06}.
In a large number of situations, LHM behave the opposite way of
conventional (right-handed) dielectric materials. This is of course the case
for refraction (so that LHM are sometimes said to 
present a negative index), but for the Goos-H\"anchen effect
as well\cite{berman02,lakh03}. Generally, lamellar structures made
with  left-handed materials present exotic properties.
It has been shown for instance that a LHM slab could support 
backward or forward guided modes\cite{shad03} and backward or
forward leaky modes \cite{wang05,moreau07,leaky}. 

In this paper we present the study of the contra-directional coupling between a guided
mode supported by a dielectric slab and a backward mode supported
by a left-handed slab\cite{shad03}. We show that this structure behaves as 
an edge-less cavity because light rotates inside the structure,
and that the modes involved in this behaviour can be excited
either using a prism coupler or by a source in one of the waveguides - 
leading to the formation of a ``light wheel''.
Such a cavity would be obtained without introducing any type of
defect, as in the case of photonic crystals.

The considered structure is presented figure \ref{f:schema} : two
slabs are surrounded by air and separated by a distance $h$. The upper
slab, whose thickness is called $h_1$, presents a relative
permittivity of $\epsilon_1$ and a relative
permeability $\mu_1$ while the lower slab, whose thickness is $h_2$,
is characterized by $\epsilon_2$ and $\mu_2$. This rather simple
structure allows analytical calculations and a rigorous analysis of
the contra-directional coupling using complex plane analysis - which
is almost never done.

We will search for solutions which present a harmonic time dependence and which do not
depend on $y$ - just like the geometry of the problem. With these 
assumptions, the solutions can be either TE polarized ({\em i.e.}
the electric field is polarized along the $y$ axis) or TM polarized
(the magnetic field being polarized along the $y$ axis). In this work,
we will only consider the TE case but the same study can be
carried out for the TM case, leading to the same conclusions.
We will hence seek a solution for which $E_y(x,z,t)= E(z)\,\exp
(i(\alpha\,x - \omega\,t))$ with $\omega=k_0\,c =
\frac{2\pi}{\lambda}\,c$ where $\lambda$ is the wavelength in the vacuum.
In the following, we will consider that $\lambda$ is the distance
unity ({\em i.e.} we take $\lambda = 1$).

\begin{figure}[h!]
\centerline{\includegraphics[width=7cm]{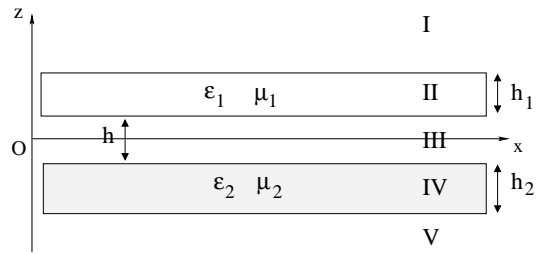}}
\caption{The two waveguides are surrounded by air and separated by a
  thickness $h$. The $z=0$ plane is placed in the middle of the air
  layer (region III). The different regions are labelled using roman numerals.
\label{f:schema}}
\end{figure}

The function $E(z)$ in a medium $j$ characterized by $\epsilon_j$ and
$\mu_j$ is in general a linear combination of two
exponential terms $\exp{(\pm i\,\gamma\,z)}$ with $\gamma_j^2 + \alpha^2 =
\epsilon_j\,\mu_j\,k_0^2$.  Since we are only interested in guided modes,
we will consider that $\alpha > k_0$ so that
the electric field is decaying exponentially in the air (we will note $\gamma_0 =
\sqrt{\alpha^2-k_0^2}$). Above and
under the waveguides (in regions I and V) only one decaying
exponential term remains. In this case, the continuity relations at
an interface between two different media can be written as a
homogeneous system of equations.
A non null solution can thus be found only if the determinant
of the system is null, which can be written
\begin{equation}\label{e:disp}
x_1\,F_1+x_2\,F_2 + (1+x_1\,x_2\,F_1\,F_2)\,\tanh (\gamma_0\,h)= 0
\end{equation}
where $F_j =
\frac{1-x_j\,\tan(\gamma_j\,h_j)}{x_j+\tan(\gamma_j\,h_j)}$ and 
$x_j=\frac{\gamma_j}{\mu_j\,\gamma_0}$ whatever the sign of
$\epsilon_j$ or $\mu_j$. Relation (\ref{e:disp}) is the
dispersion relation because each zero of the left part is a mode
supported by the structure.

The coupling of the two waveguides appears when the upper and the
lower waveguides, taken separately, both support a guided mode for
the same constant propagation, called $\alpha_0$. First we have 
chosen a small thickness for the right-handed slab ($\epsilon_1=3$,
$\mu_1=1$, $h_1=0.2\,\lambda$) so that is behaves
as a monomode waveguide and we have computed $\alpha_0$. If the two 
waveguides are identical (both right-handed and with the same
thickness) the structure supports two 
propagating modes with two close and real propagation constants.
The excitation of these two modes gives raise to an oscillation of the
energy between the two guides as shown figure\ref{f:battements}.

\begin{figure}[h!]
\centerline{\includegraphics[width=7cm]{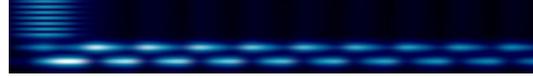}}
\caption{The two identical coupled waveguides of height $0.2\,\lambda$
  are excited using evanescent coupling : an incident beam with an incidence angle
  of $36,9$° in a medium with $\epsilon=5$ and $\mu=1$ is used. The
  energy oscillates between the two guides and leaks out when it is in
  the upper guide, so that it decays exponentially.
\label{f:battements}}
\end{figure}

When the lower waveguide is left-handed (with $\epsilon_2=-3$, $\mu_2=-1$), its thickness has to
be chosen carefully to ensure a perfect coupling with the above slab. 
Using the dispersion relation of the left-handed waveguide\cite{shad03} we have chosen the thickness of
the LHM slab ($h_2=.7146\,\lambda$) so that (i) it supports no fundamental mode (which would
otherwise propagate forward) and (ii) the only supported mode 
is more localized in the LHM and hence propagates backward.

{\em With the above parameters the dispersion relation is verified only for complex
values of $\alpha$} : the two different
solutions have then the same real part but opposite imaginary parts
({\em i.e.} conjugate propagation constants).
The solution in the complex plane are shown figure \ref{f:disp}
for different values of $h$. 

\begin{figure}[h!]
\centerline{\includegraphics[width=7cm]{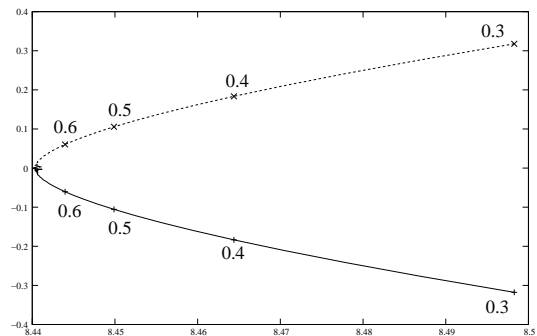}}
\caption{Solutions of the dispersion relation in the $\alpha$ complex
  plane for different values of $h$, which is given for some points
  as a fraction of the wavelength. The waveguides are characterized by
  $h_1=0.2\,\lambda$, $\epsilon_1=3$, $\mu_1=1$,
  $\epsilon_2=-3$, $\mu_2=-1$ and $h_2=.7146\,\lambda$. The units
  correspond to the choice $\lambda=1$.
\label{f:disp}}
\end{figure}

Let us now study the characteristics of the solutions, which are shown
figure \ref{f:modes}. Figure \ref{f:modes} shows the modulus, the
phase and the time-averaged Poynting vector along the $x$ direction
for the solution corresponding to the propagation
constant with a positive imaginary part with $h=0.5$.
The solution is an hybrid mode, since the field is important in both
waveguides (see figure \ref{f:modes} (a)). The field in the right-handed medium and the field in the 
left-handed medium are un phase quadrature : there is a
phase difference of $\frac{\pi}{2}$ in between as shown figure
\ref{f:modes} (b). Since the propagation constant of the other
solution is the conjugate of the one represented figure \ref{f:modes},
then the solution itself is the conjugate of the first one. The only
difference between the two solutions is hence the phase, so that
in both cases there is a phase quadrature between the two guides.
This is very different from what happens with co-directional coupled
waveguides in which the two guides are in-phase or out-of-phase,
depending on the considered mode (see figure \ref{f:battements}).

The time-averaged Poynting vector along the $x$ direction is shown figure
\ref{f:modes} (c). As expected, it is negative in the LHM and positive
elsewhere. It is very interesting to note that {\em the overall power flux of
each mode is found to be null}. This means that all the energy that is 
sent in one direction by one of the waveguides comes back using the
other one. This helps to understand the nature of the modes : although
they have an important real part of the propagation constant along the
$x$ axis, they can be considered as evanescent modes because they do
not convey any energy. It is then difficult to define a {\em propagation}
direction, but the imaginary part of the propagation constant suggests
that the modes actually have a direction in which they {\em developp}.
When the imaginary part is positive, the mode is decaying when $x$
is increasing. It is thus expected to developp towards the right - 
otherwise it would diverge. This means that the other mode developps
in the other direction. This is a discussion which is usually applied
to leaky modes\cite{leaky}.

\begin{figure}[h]
\centerline{\includegraphics[width=7cm]{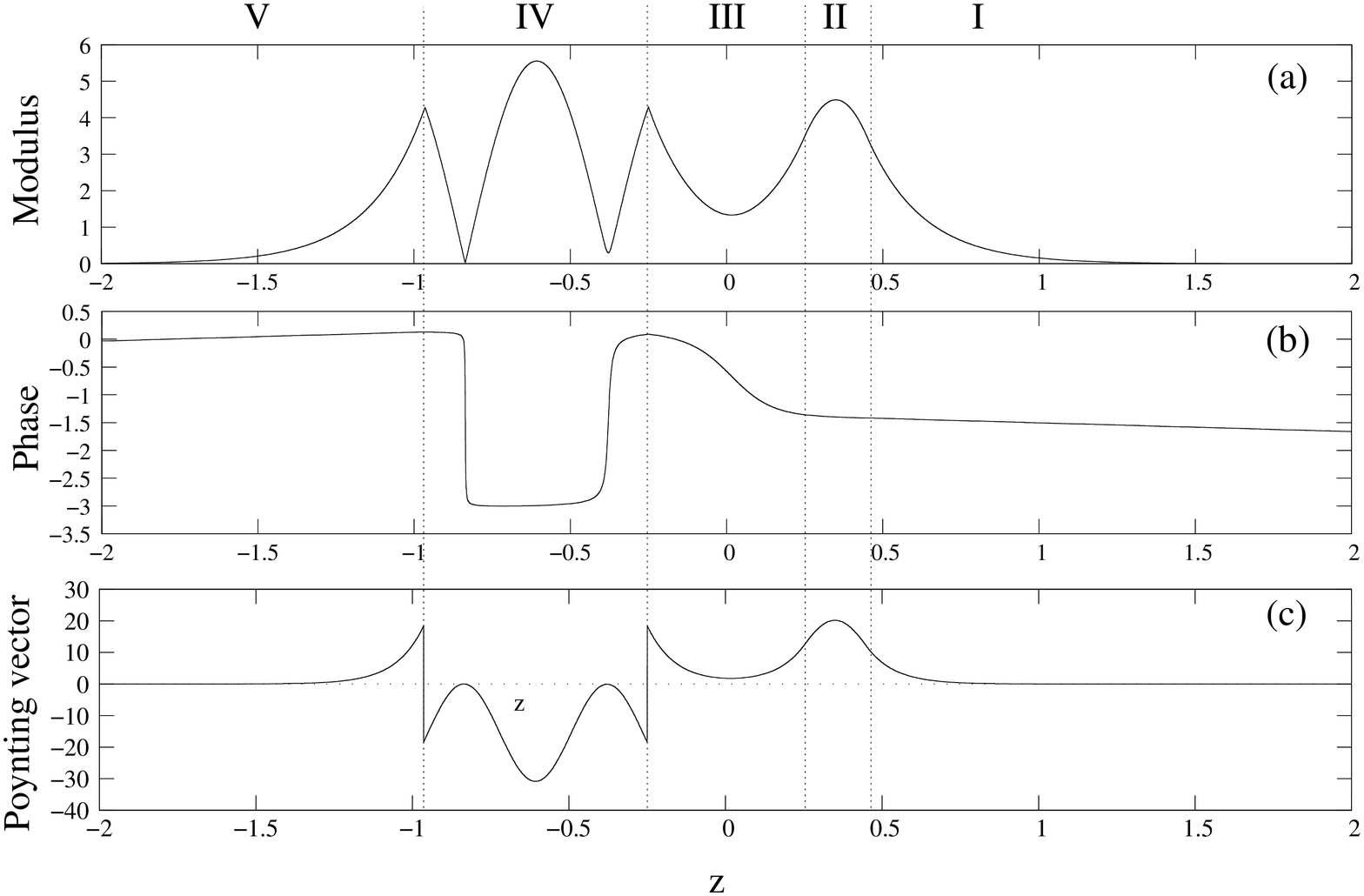}}
\caption{Characteristics of the solution with a propagation constant
  presenting a positive imaginary part for $h=0.5\,\lambda$. Figure (a) shows the
  modulus of the field (which presents two zeros in the left-handed
  guide), figure (b) the phase, which is either null or
  equal to $\pi$ in one guide and to $\frac{\pi}{2}$ in the other one,
  and figure (c) shows the Poynting vector along the $x$ direction.\label{f:modes}}
\end{figure}

Let us just underline that these two modes, having the same real part
of their propagation constant, cannot be excited separately. They can
be excited using evanescent coupling, {\em i.e.} using an incident
beam propagating in a high-index medium close to the waveguides. When
the incidence angle of the beam is greater than the critical angle,
the waveguided modes can be excited. Figure \ref{f:champ} shows such a
situation. The field is clearly enhanced in the two waveguides, and
the structure of the mode in the left-handed slab is particularly
visible. The most striking feature is the fact that there is a 
dark zone just above the incident beam in the closest slab. It is
produced by the superposition of the two modes - and the center of the
beam is the only place where it may happen since the modes develop in
different directions. Such a dark zone could be expected - after 
all, this is the case figure \ref{f:battements} - 
but in the lower waveguide. This is actually a characteristic of this
particular coupling and it persists even when the two slabs
are exchanged.

\begin{figure}[h!]
\centerline{\includegraphics[width=7cm]{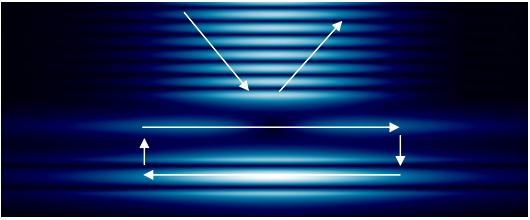}}
\caption{Excitation of the light wheel using an incident beam in a
  high-index material (with $\epsilon=5$ and $\mu=1$). The image represents the
  modulus of the $E_y$ field in a domain which is only $4.5\,\lambda$ high
  and about $60\,\lambda$ large. The propagation direction of light is
  indicated by white arrows so that the rotation of light in the
  structure is made visible. The fringes above the structure are
  localized interferences of the incident and reflected beams.
  The profiles of the modes are in perfect agreement with figure \ref{f:modes}.
  Notice the  dark zone just under the incident beam. The incident angle of the
  beam is $36.9$°, the distance between the prism and the first
  waveguide is $0.5\,\lambda$.
\label{f:champ}}
\end{figure}

Finally, when the two modes are excited they form what we called a
{\em light wheel} : 
the light is heading to the right in the right-handed waveguide and
to the left in the left-handed slab. On the right of the incident beam, the
energy flows from the right-handed slab to the left-handed slab.
This is the contrary on the left of the beam, so that light globally
rotates inside the structure. In the case of an evanescent coupling, 
the reflected beam is distorted because of the excitation of the light
wheel - but it is distorted symetrically if the coupling is
good enough. Such a simple structure could then be used to perform beam
reshaping\cite{shadrivov-beam}.

The previous results suggest that despite its lamellar geometry the structure can be
considered {\em as a cavity}. To ensure that an atom placed inside a
waveguide is actually coupled with the light wheel we have computed
the Green function of the problem. Figure \ref{f:green} shows the
Green function when a source is place in the middle of the dielectric
slab, and the light wheel is clearly excited. More precisely, two
contra-rotative light wheels are excited and interfere.

\begin{figure}[h]
\centerline{\includegraphics[width=7cm]{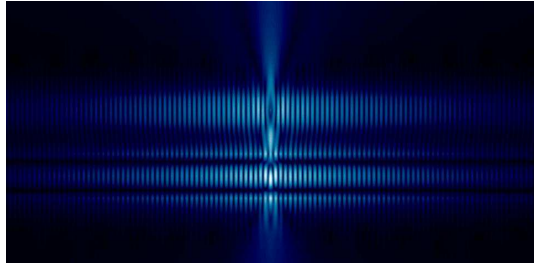}}
\caption{Modulus of the electric field when a punctual source is placed in the middle of the
  right-handed waveguide. Two contra-rotative light wheels are
  excited and they interfere which explains the interference
  fringes.\label{f:green}}
\end{figure}

In conclusion, we have shown that the use of a left-handed medium
in a lamellar structure could lead to a new type of phenomenon : 
the rotation of light which would then form a light wheel. 
This phenomenon could be used for beam reshaping, or even as a new type
of cavity. It is an original way to suppress any propagative mode in a 
dielectric slab.

The properties of the structure are likely to be improved.
Using Bragg mirrors could for instance enhance the
quality factor of such a cavity by hindering light from leaking out of the
structure. We believe that this study brings a new evidence that
left-handed materials are renewing the physics of lamellar 
structures in photonics\cite{shad05}.

The authors would like to thank D. Felbacq and E. Centeno for fruitful discussions.
This work is supported by the ANR project POEM.

\bibliographystyle{unsrt}
\bibliography{article}

\begin{thebibliography}{10}

\bibitem{veselago67}
V.G. Veselago.
\newblock The electrodynamics of substances with simultaneously negative values
  of $\epsilon$ and $\mu$.
\newblock {\em Usp.\ Fiz.\ Nauk.}, 92:517, 1967.

\bibitem{shelby01}
R.A. Shelby, D.R. Smith, and S.~Shultz.
\newblock Experimental verification of a negative index of refraction.
\newblock {\em Science}, 292:77, 2001.

\bibitem{pendry00}
J.~B. Pendry.
\newblock Negative refraction makes a perfect lens.
\newblock {\em Phys.\ Rev.\ Lett.}, 85:3966, 2000.

\bibitem{pendry06}
J.B. Pendry, D.~Schuring, and D.R. Smith.
\newblock Controlling electromagnetic fields.
\newblock {\em Science}, 312:1780, 2006.

\bibitem{berman02}
P.R. Berman.
\newblock Goos-hanchen shift in negatively refractive media.
\newblock {\em Phys. Rev. E}, 66:067603, 2002.

\bibitem{lakh03}
A.~Lakhtakia.
\newblock On plane wave remittances and goos-hanchen shifts of planar slabs
  with negative real permittivity and permeability.
\newblock {\em Electromagnetics}, 23:71, 2003.

\bibitem{wang05}
L.G. Wang and S.Y. Zhu.
\newblock Large negative lateral shifts from the kretschman-raether
  configuration with left-handed materials.
\newblock {\em Appl.\ Phys.\ Lett.}, 87:221102, 2005.

\bibitem{moreau07}
A.~Moreau and D.~Felbacq.
\newblock Comment on 'large negative lateral shifts from the kretschman-raether
  configuration with left-handed materials'.
\newblock {\em Appl.\ Phys.\ Lett.}, 90:066102, 2007.

\bibitem{leaky}
A.~Moreau and D.~Felbacq.
\newblock Leaky modes of a left-handed slab.

\bibitem{shad03}
I.V. Shadrivov, A.A. Sukhorukov, and Y.S. Kivshar.
\newblock Guided modes in negative-refractive-index waveguides.
\newblock {\em Phys.\ Rev.\ E}, 67:057602, 2003.

\bibitem{shadrivov-beam}
I.V. Shadrivov, A.A. Sukhorukov, and Y.S. Kivshar.
\newblock beam shaping by a periodic structure with negative refraction.
\newblock {\em Appl. Phys. Lett.}, 82:3820, 2003.

\bibitem{shad05}
I.V. Shadrivov, A.A. Sukhorukov, and Y.S. Kivshar.
\newblock Complete band gaps in one-dimensional left-handed periodic
  structures.
\newblock {\em Phys.\ Rev.\ Lett.}, 95:193903, 2005.

\end{thebibliography}
\end{twocolumn}
\end{document}